\documentclass{aa}

\usepackage{graphicx}
\usepackage{amsmath}
\usepackage{latexsym}

\newcommand{\gtsimeq}{\scriptsize{\raisebox{-0.7ex}{$\stackrel
        {\raisebox{-.3ex}{$\textstyle\,>$}}{\sim}\,$}}}

\begin{document}
\title{Li abundance/surface activity connections in solar-type Pleiads}
\author{D.~Clarke\inst{1} \and E.\,C.~MacDonald\inst{2} \and S.~Owens\inst{1}}
\institute{Department of Physics and Astronomy, University of
Glasgow, Glasgow, G12\,8QQ, Scotland, UK \and Department of
Astrophysics, University of Oxford, Denys Wilkinson Building,
Keble Road, Oxford, OX1\,3RH}

\offprints{d.clarke@astro.gla.ac.uk}
\date{Received <date>; Accepted <date>}
\authorrunning{D. Clarke}
\titlerunning{Activity of solar-type Pleiads}

\abstract {The relation between the lithium abundance, $A_{\rm
Li}$, and photospheric activity of solar-type Pleiads is
investigated for the first time via acquisition and analysis of
$B$ and $V$-band data. Predictions of activity levels of target
stars were made according to the $A_{\rm Li}$/$(B-V)$ relation and
then compared with new CCD photometric measurements. Six sources
behaved according to the predictions while one star (HII\,676),
with low predicted activity, exhibited the largest variability of
the study; another star (HII\,3197), with high predicted activity,
was surprisingly quiet. Two stars displayed non-periodic fadings,
this being symptomatic of orbiting disk-like structures with
irregular density distributions. Although the observation windows
were not ideal for rotational period detection, some periodograms
provided possible values; the light-curve obtained for HII\,1532
is consistent with that previously recorded.
\keywords{ stars: solar-type  -- stars: rotation } }

\maketitle

\section{Introduction}
The observational study by Duncan \& Jones (\cite{DJ}) of G \& K
type stars in the Pleiades indicated a much larger dispersion of
surface lithium abundance, $A_{\rm Li}$ [$= 12 + \log \{N({\rm
Li})/N({\rm H})\}]$, than expected for a cluster of its age
($\sim\,$75/100\,Myrs). An early explanation suggested that the
members form an age dispersion (Duncan \& Jones \cite{DJ}; Butler
et al. \cite{Betal}) but this notion has since been abandoned.
Subsequent studies by Soderblom et al. (\cite {Setal1}) and Jones
et al. (\cite{Jetal}) confirmed the $A_{\rm Li}$ dispersion, being
most pronounced for stars with $M < 0.9\,M_{\odot}$. They claimed
that rapid rotators (high $v\sin i$) have preserved more of their
initial Li relative to slow rotators of similar mass. Some fast
rotators appear to have preserved all their initial Li. According
to Jones et al. (\cite{Jetal}), the association of high $v\sin i$
with undepleted $A_{\rm Li}$ results from rapid rotation affecting
the circulation current sufficiently to prevent Li accumulating
near the bases of the convective zones where the temperature is
sufficient for Li to be destroyed. The observational studies of
Soderblom et al. (\cite {Setal1}) and Jones et al. (\cite{Jetal})
also showed that stars with low Li-depletion have a tendency to
display chromospheric activity, although this alone was unable to
account for the $A_{\rm Li}$ dispersion.

Concurrently, Soderblom et al. (\cite{Setal2}) demonstrated that
chromospheric activity, based on emission measures of H$\alpha$
and Ca\,II\,8542\,\AA\,, was correlated with rotation after
normalising $v\sin i$ values to a parameter akin to the Rossby
number of each star. Many stars display temporal variability in
the chromospheric lines, degrading the correlation, based as it is
on synoptic measurements. The question of the dispersion of
$A_{\rm Li}$ at a given stellar mass being related to temporal
variability in the Li\,I 6708\,\AA\ line has, been ruled out,
however, by Jeffries (\cite{Jeffries}).

In pursuit of understanding the reasons for the $A_{\rm Li}$
dispersion, King et al. (\cite{Kingetal}) have re-examined many of
the complex issues of the Li/rotation connection in more
comprehensive ways. They have presented evidence that dispersion
in equivalent widths of the Li line is caused by stellar
atmospheric effects rather than being wholly related to genuine
abundance differences. Using more rigorous statistical methods
than in previous studies, they have confirmed a strong correlation
between stellar activity and Li-excess but not a one-to-one
perfect mapping between $A_{\rm Li}$ and stellar rotation, based
on determined periods rather than $v\sin i$ projections.

All the discussions within the papers cited above bear testament
to the very complex issues relating the various observed
parameters and to the problems addressed to explain the range of
$A_{\rm Li}$ in coeval stars, this being key to our understanding
of early stellar evolutionary processes and of their time scales.

Recently, Messina et al. (\cite{Messetal}) have investigated the
``rotation-photospheric activity connection" using the diagnostics
of $V$-band light-curve amplitudes as being indicators of starspot
coverage and magnetic activity. From a compilation of various data
they have found a high degree of correlation between the envelope
of maximum $V$-band light-curve amplitudes and the rotation period
in five open clusters, including the Pleiades. In this paper the
relationship between $A_{\rm Li}$ and {\it photospheric} activity
is explored for the first time by assessing data from the
literature and by new measurements presented here.

Figure~1, provides the $(B-V)/A_{\rm Li}$ diagram for the Pleiades
with additional information on photospheric activity and rotation.
The basic positions of each star were obtained from Jones et al.
(\cite{Jetal}) and Soderblom et al. (\cite{Setal1}). Activity
estimates were obtained from broad-band photometry for 27 stars
from van Leeuwen \& Alphenaar (\cite{VanL1}), Panov \& Geyer
(\cite{Pan}), Magnitskii (\cite{Mag1}, \cite{Mag2}), Stauffer \&
Hartmann (\cite{Stauff}) and O'Dell et al. (\cite{ODell}); the
estimates are on a sliding scale with high activity set at
$\gtsimeq\,0\fm25$, medium activity $\sim 0\fm1$, and low activity
being just above detectable limits $(\sim 0\fm01)$. These activity
estimates have been combined with rotational velocity data from
Stauffer \& Hartmann (\cite{Stauff}). Inspection of Fig.~3 in
Stauffer et al. (\cite{Stauffetal}), suggests a bimodal
distribution of rotational velocities with fast rotators being in
a group with $v\sin i\,\gtsimeq 30\,$km\,s$^{-1}$; stars with
$v\sin i\,>30\,$km\,s$^{-1}$ have been classed as ultra-fast
rotators (UFRs) by Soderblom et al. (\cite{Setal2}).

The stars with estimated activities are plotted for UFRs, with
increasing sizes of diamond [$\diamond\longrightarrow\Diamond$],
according to the amplitude of activity; for slow rotators $(v\sin
i<30\,$km s$^{-1})$, the degree of activity is noted by increasing
sizes of circle [$\circ\longrightarrow$ {\raisebox{-1pt}{O}].
Stars of unknown activity are simply marked as `$\mathbf{ +}$' or
`$\mathbf{\times}$' according to whether they are slow or rapid
rotators respectively. The spread in $A_{\rm Li}$ is clear for
stars of $(B-V)>0.7$ (late G- to K-type). For stars of a given
colour, the dispersion reveals a tendency for the more
photospheric active stars to have higher $A_{\rm Li}$. The same
tendency related to fast rotators, referred to earlier, can also
be seen.

\begin{figure}
\centering
\caption {The $(B-V)/A_{\rm Li}$ diagram for the Pleiades with 27
stars, monitored previously for photospheric activity,
highlighted. For fast rotators $(v\,\sin i > 30$\,km\,s$^{-1})$,
diamond symbols [$\diamond\longrightarrow\Diamond$] indicate the
degree of activity by their size; for slow rotators $(v\sin i <
30$\,km\,s$^{-1})$, increasing sizes of circle
[$\circ\longrightarrow$\,O] indicate the degree of activity. The 8
target stars are indicated by their HzC number. Another marked
star, HII\,738, in the same field of a target and previously
measured for activity, was also monitored. Other stars are simply
marked according to their rotation being rapid
(`${\mathbf\times}$') or slow (`${\mathbf{+}}$').} \label{F:1}
\end{figure}

In order to extend the data base for degree of variability of
stars with measured $A_{\rm Li}$ and to gain better understanding
of the $A_{\rm Li}$/activity tendency, 8 stars marked in Fig.~1 by
catalogue number (Hertzsprung \cite{Hertz}) [later referred to as
HzC] were observed with the JKT in La Palma using the CCD imager
(TEK4 detector); some 40 other stars within the target fields were
investigated. Data for $B$ and $V$ filters were obtained during
November 1998, the 7 consecutive nights having Julian Dates
running from 245\,1141.3 to 245\,1147.7, each night simply being
referred to as JD41 to JD47 inclusively. From the suggested
tendency highlighted above, the activities of the 8 targets were
predicted according to their position in the $(B-V)/A_{\rm Li}$
diagram, with details given in Table~1, towards the end of the
paper; $B$ and ($B-V$) are taken from photoelectric measurements
of Johnson \& Mitchell (\cite{JM}) [later referred to as J\&M].
The major aim of the exercise was to compare these predicted
activity estimates with the new photometric observations.

\section{Data Reduction and Analysis Techniques}
The targets were positioned in the frame to allow acquisition of a
selection of field stars required for differential photometry.
Where possible, $B$ and $V$ frames for each target were taken as
sequence pairs with exposures from 10\,s to 120\,s according to
the brightness of the target and the seeing quality. Dark, bias
and twilight flat-field frames were obtained as standard
procedure. Data reduction was performed using the Starlink package
CCDPACK. Once the frames were flat-fielded, bias subtracted and
aligned, the package AUTOPHOTOM was implemented to extract
magnitudes for the monitored stars. This reduction gives error
estimates based on the variance component of the data array which,
for most frames, gave determinations correct to $\sim \pm
0\fm001$, with differential photometry achieved to an accuracy of
a few milli-mags.

Although each field contained at least one star with $B\,,V$
values from J\&M, no true standards were observed. Without a full
transformation procedure, the values reported here are essentially
``instrumental" and designated $B_i\,,V_i$, although any
differences from their underlying standard $B\,,V$ values are
likely to be very small. In considering the results, it must be
remembered that the $B\,,V$ reference values of J\&M are synoptic
and may carry systematic errors since these stars are variable.
Further small systematic errors may also be present in the mean
$B_i\,,V_i$ of each field star resulting from their own possible
variability. In assessing the variability of each monitored star,
a problem arises from not knowing which field star can be taken as
a stable reference. Since the stars have similar brightness, they
are likely to be of similar spectral type and are all probably
variable to some degree.

Firstly, the problem was tackled by taking each star in turn to
act as reference, so providing magnitude differences designated by
$\Delta B_i\,,\Delta V_i$, according to the passband. Assessment
of activity was made by inspection of these values simply plotted
as time sequences. An example from such an exercise is given in
Fig.~\ref{F:2}. Inter-comparison of all the various temporal
patterns of $\Delta B_i\,,\Delta V_i$ provided insight on the
activity of each star. If, for example, relative to a particular
reference star, the temporal variations $\Delta B_i\,,\Delta V_i$
for all the other stars displayed the same signatures, this
indicated variability of the reference object itself. Selection of
the best comparison stars was made by considering magnitude
differences which were not mirrored from one star to the next, or
which for some star pairs, showed no statistically significant
variations at all.

\begin{figure}
\centering
\caption {Using HII\,229 as the reference star, $\Delta B_i$
values for HII\,253 show night-to-night variations, with drifting
changes on each night particularly well seen around JD\,44, at the
phase of the light-curve mid-point when the changes are at their
fastest. For HII\,228 there is a steady rise in brightness over
the 6 nights, probably related to a longer period of variability.}
\label{F:2}
\end{figure}

Secondly, for those occasions when $B\,,V$ frames were exposed in
quick succession, the contemporaneous differences, $\Delta
B_i\,,\Delta V_i$, may be plotted against each other. According to
van Leeuwen et al. (\cite{VanL}), periodic variable solar-type
Pleiads display only marginal colour index changes. $B$ and
$V$-band variations are therefore correlated, and display a linear
relationship. If the disturbing features, with small temperature
differences relative to the undisturbed photosphere, traverse the
projected disk, the change in $B$ is slightly greater than for $V$
and hence the $B/V$ gradient should be just greater than unity;
this is independent of whether the features are bright patches or
dark spots. The determined gradient is independent of any
interstellar extinction affecting the star's colour. Whether or
not stars are active can therefore be checked by investigating the
correlation between the $\Delta B_i\,,\Delta V_i$ values. As the
measurements were made under fairly constant conditions, with very
similar error values for the differential magnitude
determinations, the required $\Delta B_i/\Delta V_i$ and $\Delta
V_i/\Delta B_i$ gradients for the exercise were calculated without
weighting of the individual measurements, with a correlation
coefficient determined by the standard method. Following any
marked correlation, the degree of activity was estimated from the
spread of the $\Delta V_i$ values in the $\Delta B_i\,,\Delta V_i$
plot. For data with a low level of correlation but with
dispersions of $\Delta B_i\,,\Delta V_i$ larger than that
associated with measurement noise, both stars may be considered as
being variable. For star pairs with little or no activity, the
$\Delta B_i\,,\Delta V_i$ values overlap within a spread of their
joint distribution according to the measurement uncertainties.

Two examples of the behaviour of contemporaneous $\Delta
B_i\,,\Delta V_i$ measurements are given in Fig.~\ref{F:3}, one
displaying strong activity of either the target star or its
reference, the other showing little or no variation of the two
compared stars to the limit of the measurement noise.

\begin{figure}
\centering
\caption{(a) Contemporaneous $\Delta B_i\,,\Delta V_i$ magnitude
differences of HII\,676\,$-$\,HII\,655 are displayed, indicating a
large variability of the target star over a range $\sim 0\fm5$.
(b) The $\Delta B_i\,,\Delta V_i$ behaviour of
HII\,711\,$-$\,HII\,655 suggests that neither star displays any
significant variability greater than the noise $\sim \pm 0\fm01$.}
\label{F:3} \vspace{-5mm}
\end{figure}

The data distribution in the $\Delta B_i\,,\Delta V_i$ plane may
also provide insight on the variability. If a star is measured at
random and has a sinusoidal-like variation, the values will be
located according to the principles of simple harmonic motion,
with more measurements likely to be made at the light-curve maxima
and minima, where the rates of change are slowest. The density of
points will tend to concentrate at the extremes of the linear
distribution rather than at mid-level values; such behaviour shows
in Fig.~\ref{F:3}(a).

Thirdly, the data were subjected to string/rope analysis,
providing periodograms from the regularised Lafler-Kinman (LK)
statistic, $T(P)$, (see Clarke \cite{Clarke}: Eq.~(3)).
Periodograms were investigated for the $B$ and $V$-bands
separately and their combination. As several stars were monitored
in each recorded frame, those stars deemed to be stable were used
to explore sampling and windowing effects. Because of the
relatively short time span of this photometric exercise, it was
not expected that periods would be determinable for most of the
monitored stars. A good example of the behaviour of an LK
periodogram and a resulting light-curve are provided in
Fig.~\ref{F:4} for HII\,1532.

\begin{figure}
\centering
\caption{(a) The LK periodograms based on $\Delta B_i\,,\Delta
V_i$ for HII\,1532\,$-$\,HII\,1575 provide a minimum at $0\fd777$.
\ (b) The $\Delta B_i\,,\Delta V_i$ values phased on a period of
$0\fd777$ show the outlines of the underlying coherent
light-curves.} \label{F:4}
\end{figure}

The three analysis schemes outlined above were applied to each of
the 8 selected fields. The monitored stars are identified in
Table~1 with their determined values of $B_i\,,V_i$ and
$(B_i-V_i)$.

\section{The Results}
\subsection{Field 1: HII\,253}
{\it Summary}: $B$-band 60 frames; $V$-band 50 frames.\\
Six nights of observation from JD\,42 to JD\,47.\\
J\&M lists only HII\,263 for reference $B,V$ values.\\
Six field stars monitored with HzC identification.\\
HII\,219 was taken as the stable, reference object.

The temporal behaviour of HII\,253 (see Fig.~\ref{F:2}) gives
evidence of an underlying period. On JD\,44, the brightness
displayed a progressive drop possibly at the falling mid-point of
a light-curve when the changes are at their fastest. The range of
the variation is $\sim 0\fm2$. The lower portion of Fig.~\ref{F:2}
reveals HII\,228 to be variable, with a smaller range ($\sim
0\fm08$) and with a period longer than 7\,days.

The $\Delta B_i\,,\Delta V_i$ plot based on
HII\,253\,$-$\,HII\,219 confirmed the target's variability with
data density enhancement at minimum brightness, suggesting the
sampling of a sinusoidal-like light-curve. The paucity of
measurements at higher brightness suggest the observation windows
do not cover the light maximum. Although $\Delta V_i$ had a range
$\sim 0\fm16$, the true span is probably greater, perhaps $\sim
0\fm2$, giving a light-curve amplitude $\sim 0\fm1$. The
determined $\Delta B_i/\Delta V_i$ gradient was 1.06, a value in
keeping with disturbed patches on the stellar surface with
temperatures differing by a few hundred degrees from the mean
photosphere, the same cells also causing the brightness changes.

The periodogram for $\Delta B_i$ gave minima at $1\fd365$ and
$1\fd717$. The former period provides a near sinusoidal
light-curve but with a sharp departure over a phase interval $\sim
0.1$ (see Fig.~\ref{F:5}(a)). This ``spike" relates to data
recorded on the last night. It also occurs in the $V$-band
light-curve and is likely caused by a rapid change in the
disposition of the surface features rather than by flare activity.
Marilli et al. (\cite{Marilli}) have reported a period of
$1\fd721$. If the data are phased on the suggested close value of
$1\fd717$, the ensuing light-curve (Fig.~\ref{F:5}(b)) shows a
more complicated structure. Further observations are required to
confirm the period of this star.

\begin{figure}
\centering
\caption{(a) Phased on a period of $1\fd365$ the $\Delta B_i$ for
HII\,253\,$-$\,HII\,219 provides a sketched near-sinusoidal
light-curve but with spike. \ (b) The same data phased on a period
of $1\fd717$ provides the outline of a less distinctive
light-curve.} \label{F:5}
\end{figure}

HII\,272 was variable with changes of $\sim 0\fm05$ but with a
weak correlation between the $B$ and $V$ measurements, the result
perhaps being influenced by possible small scale variability of
the reference star, HII\,219. The data for both HII\,203 and
HII\,217, set variation limits of $\sim 0\fm01$. However, the
important outcome of this exercise is the scale of the light
variations of HII\,253, confirming the medium/high level activity
prediction in Table~1.

Being redder than other field stars and with brightnesses just
above the main sequence trend (see Fig.~\ref{F:9}), HII\,217 and
HII\,272 are candidates for possessing unresolved companions. The
population of such solar-type photometric binaries within the
Pleiades is estimated at 26\% by Stauffer (\cite{Stauff1}).

\subsection{Field 2: HII\,263}
{\it Summary}: $B$-band 41 frames; $V$-band 34 frames.\\
Six nights of observation.\\
J\&M lists only HII\,263 for reference $B,V$ values.\\
Five field stars with HzC identification.\\
HII\,257 displayed least activity --- used as stable reference.\\
HII\,282 showed discrepant brightness compared with HzC.

The $\Delta B_i\,,\Delta V_i$ values were strongly correlated with
a gradient of 1.16 and with $\Delta V_i$ varying by $\sim 0\fm18$,
validating the prediction for HII\,263 in Table~1.

A suggested period of $0\fd851$ produced in-phase rudimentary
light-curves for the two colours (see Fig.~\ref{F:6}(a)). Based on
HII\,263$\,-\,$HII\,322 and HII\,263$\,-\,$HII\,309, periods of
$0\fd859$ and $0\fd848$ also emerged. According to Krishnamurthi
et al. (\cite{Ketal}), HII\,263 has a period of $4\fd82$; folding
the data on this covered only three quarters of the cycle but with
fluctuations, suggesting a poor fit. A longer data run and a
further investigation of reference stars are required to confirm
the correct period.

HII\,282 displayed a peculiar behaviour. For most of the run,
$\Delta B_i/\Delta V_i$ data gave a strong correlation with a
gradient of 1.17, similar to the behaviour of other spotted
rotating stars. On the night of JD\,45, however, the star faded
with a fall $\sim 0\fm3$ before recovering, the episode covering
$\sim 3\fh2$ (see Fig.~{\ref{F:6}(b)). The three measurements on
JD\,42 also provide anomalous low brightness values. As such
behaviour was unexpected, particular checks were made on these
data frames. There was no evidence of poor image registrations;
the anomalies only affected HII\,282 and, as it was fairly central
to the field, no technical reason was apparent to explain the
behaviour. Its $m_{pg}$ value in HzC differs by $\sim 5$ mags from
the mean $B_i$ determined here. Thus the HzC value may have been
obtained either at a time of extreme low brightness or there is a
misprint in the catalogue. Its position relative to the main
sequence trend (see Fig.~\ref{F:9}) and its red colour with
respect to the other field stars make it a possible photometric
binary candidate, raising a question of the system being a
short-period eclipsing binary involving a faint red dwarf. The
behaviour of the events are more akin, however, to the
non-periodic Algol-type minima associated with HAEBE objects, with
UX\,Ori as prototype (Bibo \& Th\'e, \cite{BT}). For these young
objects, the photometric behaviour is modelled in terms of
irregular dust flows within disk-like envelopes, producing
variable scattering and extinction (see, for example, Rostopchina,
et al. \cite{Rosto}). HII\,282 certainly deserves a more intensive
photometric investigation.

\begin{figure}
\centering
\caption{ (a) The $\Delta B_i\,,\Delta V_i$ differences for
HII\,263\,$-$\,HII\,257 phased on a period of $0\fd848$ show a
rudimentary light-curve, the two-colour variations being in phase.
(b) The $\Delta B_i\,,\Delta V_i$ correlation of
HII\,282\,$-$\,HII\,257 shows fading of HII\,282 $\sim 0\fm3$ on
JD\,42 and JD\,45; on the latter night, the star faded and
recovered, the episode covering $\sim 3\fh2$.} \label{F:6}
\end{figure}

The $\Delta B_i\,,\Delta V_i$ values for HII\,322\,$-$\,HII\,257
also indicate a strong variability of HII\,322 with range $\Delta
V_i \sim 0\fm16$. Slightly larger variability was noted for
HII\,309 with $\Delta V_i \sim0\fm18$. Periodograms for HII\,322
and HII\,309 both displayed similar structures likely caused by
sampling and windowing effects and no periods could be assigned.

\subsection{Field 3: HII\,522}
{\it Summary}: $B$-band 47 frames; $V$-band 38 frames.\\
J\&M lists only HII\,522 for reference $B,V$ values.\\
Number of nights monitored: 6.\\
3 field stars monitored with HzC identification.\\
HII\,575 displayed least activity --- used as stable reference.

The activity of the monitored stars was noticeably less than for
the two previously discussed fields. HII\,522 was the most active
star. Intercomparison of the various time plots revealed
brightness changes on the nights of JD\,45 and JD\,46; on other
nights any variability was marginal. No periodicities were
detectable in any of the stars.

The $\Delta B_i\,,\Delta V_i$ plots of HII\,522 showed most values
contained within a small area appropriate to the measurement
uncertainties, indicating low activity. Between HJD 46.483 to
46.696 it became $\sim 0\fm1$ fainter, however, but with no
significant colour change. This signature could not be ascribed to
any photometric technical problem. Its position relative to the
main sequence trend (see Fig.~\ref{F:9}) makes it a photometric
binary candidate. The fading behaviour was similar to HII\,282 and
again might be ascribed either to the star being a short-period
eclipsing binary or to irregular orbiting dust clouds imposing
their effects on a small rotational modulation of active
photospheric areas, the scale of the latter supporting the
prediction of Table~1.

It may be noted also that on the nights JD\,41 to JD\,43 no
variations between HII\,575 and HII\,598 were detectable, as also
for the nights JD\,44 to JD\,47. If the two time sections are
averaged, however, HII\,598 increased in brightness by $\sim
0\fm05$ relative to HII\,575.

All the colour values are larger than expected, typically by
$0\fm3$. A polarimetric study by Breger (\cite{Breger}) revealed
that this cluster area is affected by interstellar material. One
of the field stars, HII\,575, classed as a cluster non-member,
displays $p = 0.93\%$, a value sufficiently high to account for
its apparent colour excess.

\subsection{Field 4: HII\,676}
{\it Summary}: $B$-band 33 frames; $V$-band 34 frames.\\
Observations on nights JD\,44, JD\,45, and JD\,46 only.\\
4 field stars monitored with HzC identification.\\
J\&M lists HII\,625, HII\,676 and HII\,738 with $B,V$ values.\\
HII\,738 used as the magnitude reference.\\
HII\,655 displayed least activity --- used as stable reference.

Although noted by Krishnamurthi et al., (\cite{Ketal}) as
periodic, HII\,738 exhibited smaller activity than HII\,625 and
HII\,676 and was used as reference to provide mean $B_i\,,V_i$
magnitudes for the other stars. Comparison with J\&M shows
HII\,625 to be fainter by $0\fm09$ and $0\fm07$ in $B_i$ and $V_i$
respectively, while HII\,676 is brighter by $0\fm14$ and $0\fm21$
for the two bands. Such differences highlight the earlier
discussed problems of assigning magnitudes in this exercise.

The $\Delta B_i\,,\Delta V_i$ plot for HII\,711\,$-$\,HII\,655
(see Fig.~\ref{F:3}(b)) shows neither star exhibiting variability;
the tightly distributed data have dispersion according to the
measurement noise $(\sim \pm 0\fm007)$. Using HII\,655 as
reference, the target star (HII\,676) and HII\,625 both displayed
high activity $\sim 0\fm5$. With respect to the prediction,
HII\,676 is more active than expected.

HII\,738 shows $\Delta V_i$ to vary by $\sim 0\fm15$. Listed in
Soderblom et al. (\cite{Setal1}), with an $A_{\rm Li}$ value of
2.6, it would be expected to display medium to high activity, the
measurements here agreeing with the prediction.

From three nights only, period detections were not anticipated.
Alternative periods for HII\,738 are noted in the literature:
Krishnamurthi et al. (\cite{Ketal}) give $0\fd83$ while Marilli et
al. (\cite{Marilli}) give $1\fd460$. This discrepancy could not be
resolved by the limited data here. Although the periodogram for
HII\,625 suggested a period of $0\fd466$, close to the values of
$0\fd422$, determined by Magnitskii (\cite {Mag1}) and $0\fd428$
by van Leeuwen et al. (\cite{VanL}), the LK minimum was not
statistically significant. More extensive photometry is certainly
required for this field.

All the stars have larger colour values than expected with
excesses just greater than $0\fm3$. Breger (\cite{Breger}) showed
this area to display high interstellar polarization and
significant reddening. Stauffer \& Hartmann (\cite{Stauff}) give
reddening estimates of 0.34, 0.28 and 0.40 for HII\,625, HII\,676
and HII\,738 respectively. It may be noted that HII\,738 has been
found by Bouvier et al. (\cite{Bouvier}) to have a companion
separated by $0\farcs50$ and this may affect the perceived colour.

\subsection{Field 5: HII\,1532}
{\it Summary}: $B$-band 38 frames; $V$-band 37 frames.\\
Observations on 4 nights JD\,44 to JD\,47.\\
J\&M lists only HII\,1532 with $B,V$ values.\\
10 field stars monitored, 9 with HzC identification.\\
HII\,1575 used as stable reference.

HII\,1532 was markedly the most variable in the field. Periods
from $0\fd771$ to $0\fd795$ were obtained, according to the chosen
reference star. Based on $\Delta B_i\,,\Delta V_i$, joint
periodograms from HII\,1532\,$-$\,H\,II\,1575 gave $0\fd777$ with
coherent light-curves for the two bands as depicted in
Fig.~\ref{F:4}(b). Krishnamurthi et al. (\cite{Ketal}) reported a
period of $0\fd78$, with variation of similar form. A brightness
enhancement, covering about half of the period, suggests the
presence of a single feature traversing the projected stellar disk
and being obscured during the other half of the cycle. The range
of the variation of $V_i$ was $\sim 0\fm15$, in keeping with the
Table~1 prediction.

HII\,1582 displayed small activity. Its periodogram suggested a
period $\sim 0\fd95$ but the engendered light-curves had
inadequate phase coverage, being based on only four consecutive
nights of observation. The variability was small, $\sim 0\fm05$ in
the V-band. Little or no variability was detected for the other
field stars.

Four of the listed stars in Table~1 have colour values $>1\fm0$
suggesting that they suffer from reddening. Whether this results
from interstellar material or more localised dust might perhaps be
resolved by further photometric studies using a larger range of
wavebands or by polarimetric studies. The field is towards the
cluster edge and was beyond the areas covered by Breger's
(\cite{Breger}) polarimetric study and was not included in the
reddening estimates of Stauffer \& Hartmann (\cite{Stauff}).

\subsection{Field 6: HII\,1553}
{\it Summary}: $B$-band 28 frames; $V$-band 28 frames.\\
Observations on nights JD\,42, JD\,43, and JD\,44 only.\\
J\&M lists only HII\,1553 with $B,V$ values.\\
4 field stars monitored, 2 with HzC identification.\\
HII\,1508 used as stable reference.

All the monitored stars appeared to be relatively inactive. Only
HII\,1553 provided a recognisable signature showing a maximum on
JD\,43 and a progressive fall on JD\,44, suggesting a period of
just less than one day. The sparse data produced periodograms with
no statistically significant minima. The variation in $\Delta V_i$
was $\sim 0\fm07$, this being a little below the prediction for
HII\,1553 in Table~1.

For the other four stars, all are variables of small amplitude.
The $\Delta B_i\,,\Delta V_i$ plot for HII\,1554$\,-\,$HII\,1508,
showed no strong correlation but the dispersion was in excess
expected from the measurement uncertainties, the pattern
suggesting that both stars vary in $V_i$ by $\sim 0\fm03$.

HII\,1553 appears redder (listed in J\&M) relative to the field
stars. From multi-colour photometry by Stauffer (\cite{Stauff1}),
it has been classed as a photometric binary this being confirmed
by Bouvier et al. (\cite{Bouvier}), the companion separated by
$0\farcs09$.

\subsection{Field 7: HII\,1776}
{\it Summary}: $B$-band 14 frames; $V$-band 13 frames.\\
Observations on nights JD\,41, JD\,42 and JD\,46 only.\\
J\&M lists only HII\,1776 with $B,V$ values.\\
4 field stars monitored, 2 with HzC identification.

From the limited run, it was difficult to decipher the behaviour
of the monitored stars. Cursory inspection suggested that
HII\,1776 is variable but not to the degree of HII\,1844. On
JD\,42, this latter star gradually brightened by $\sim 0\fm1$
relative to both HII\,1776 and HII\,1900 and may be described as
displaying medium level activity.

HII\,1844 displayed an anomalous $B$ magnitude (see
Fig.~\ref{F:8}) possibly as a result of flare activity. HII\,1900
is redder in relation to other field stars, lying above the
general trend of the main sequence (see Fig.~\ref{F:9}), and is
probably a photometric binary.

\subsection{Field 8: HII\,3197}
{\it Summary}: $B$-band 55 frames; $V$-band 55 frames.\\
Observations on JD41, JD42, JD44, JD46 and JD47.\\
J\&M lists only HII\,3197 with $B,V$ values.\\
4 field stars monitored, 2 with HzC identification.\\
\#7 used as stable reference.

The stars labelled \#6 and \#7 exhibited no detectable
variability. Surprisingly, no variability was detected for
HII\,3197 above the measurement noise. Any variability was less
than $\sim 0\fm03$. Krishnamurthi et al. (\cite{Ketal}) give a
period of $0\fd44$ with amplitude $\sim 0\fm03$. From the
distribution of $\Delta B_i\,,\Delta V_i$ values based on
HII\,3197\,$-$\,\#7, period detection was not expected, this
proving to be the case.

\begin{figure}
\centering
\caption {The $\Delta B_i\,,\Delta V_i$ light-curves of HII\,3197
based on a period of $0\fd901$ using values from
HII\,3167\,$-$\#7.} \label{F:7}
\end{figure}

HII\,3167 was the most active of stars in the field. The $\Delta
B_i\,,\Delta V_i$ plot revealed variability of $\sim 0\fm25$ and
$\sim 0\fm15$ in the respective bands. A suggested period of
$0\fd901$ emerged, although one at $\sim 1\fd4$ was also possible.
Based on the former, the phased data are displayed in
Fig.~\ref{F:7}, the amplitude for $\Delta B_i$ being greater than
for $\Delta V_i$. Further data are required to confirm the period.
HII\,3198 also displayed variability with a range $\sim 0\fm08$.

The colour index for HII\,3197 listed in J\&M is anomalous
relative to the other field stars. Stauffer (\cite{Stauff1}) has
classed it as a photometric binary; Bouvier et al.
(\cite{Bouvier}) have resolved the system as being triple.

\begin{table*}
\caption{The monitored stars in 8 fields are tabulated with their
mean $B_i$ magnitudes and colour values, $B_i - V_i$. At the head
of each block, the target star is noted with indication of its
rotation speed ([F]ast or [S]low) and its activity prediction (1 =
high, 2 = medium and 3 = low). The star taken as the field
magnitude reference is also given, its $B_i$ and $B_i - V_i$
values corresponding to $B$ and $B - V$ in J\&M. Under ``Var
$V_i$" estimates are noted of the total $V_i$ variation due to
photospheric features. The co-ordinates ascribed to the
non-cataloged stars listed in the bottom left hand section are for
Epoch 2000.0 and are estimates based on the other stars in the
field and on the plate scale.}
\begin{center}
\begin{tabular}{l|cccl||l|cccl}
\hline HII \ \ $m_{pg}$\ \ $A_{\rm Li}$
&$B_i$&$B_i-V_i$&Var\,$V_i$&Comments&
HII \ \ $m_{pg}$\ \ $A_{\rm Li}$ &$B_i$&$B_i-V_i$&Var\,$V_i$&Comments\\
\hline
\multicolumn{1}{c|}{\bf Field 1}&\multicolumn{4}{l||}{{\bf HII\,253 [F: 2/1]} \quad {\bf Ref: HII\,253}}&
\multicolumn{1}{c|}{\bf Field 5}&\multicolumn{4}{l}{{\bf HII\,1532 [S: 3/2]}\quad {\bf Ref: HII\,1532}}\\
203 15.13 \ \ ---&15.26&$+0.72$&$< 0\fm01$&\qquad --- &1427 14.70 \ ---& 15.05 &$+1.22$& --- &\qquad --- \\
217 13.11 \ \ ---&12.90&$+1.07$&$< 0\fm01$&\qquad --- &1457 15.41 \ ---& 15.90 &$+0.66$& --- &\qquad --- \\
219 14.71 \ \ ---&14.78&$+0.74$& --- &\qquad --- &\#1\quad\ \ ---\quad\ ---  & 17.57 &$+0.79$& --- &\qquad --- \\
223 16.1 \quad\,---&16.37&$+0.80$& --- &\qquad --- &1478 15.92 \ ---& 16.51 &$+1.05$& --- &\qquad --- \\
228 12.57 \ \ ---&12.38&$+0.61$&$0\fm08$&\qquad --- &1493 16.2  \ \, ---& 16.41 &$+0.63$& --- &\qquad --- \\
253 11.30 \ \ 3.13&11.34&$+0.68$&$0\fm20$&P\,$=1\fd365$   &1532 14.92 \ 0.37&15.21&$+1.26$& $0\fm15$&P\,$=0\fd777^*$\\
272 14.31 \ \ ---&14.09&$+1.70$&$0\fm05$      &Very red  &1534 16.4  \ \, ---& 16.79 &$+0.80$& --- &\qquad --- \\
\cline{1-5} \multicolumn{1}{c|}{\bf Field 2}&\multicolumn{4}{l||}{{\bf HII\,263 [S: 1]}\quad {\bf Ref: HII\,263}}&
1545 15.39 \ ---&15.81&$+0.72$& --- &\qquad --- \\
225 13.97 \ \ ---&14.34&$+0.73$& --- &\qquad --- &1570 14.89 \ ---& 15.13 &$+1.08$& --- &\qquad --- \\
257 13.30 \ \ ---&13.50&$+0.84$& --- &\qquad --- &1575 16.6  \ \, ---& 16.94 &$+0.87$& --- &\qquad --- \\
263 12.30 \ \ 3.11&12.42&$+0.88$&$0\fm18$&P\,$=0\fd851$   &1582 15.42 \ ---  & 15.66 &$+1.27$& $0\fm05$&\qquad --- \\
\cline{6-10} 282 15.90 \ \ ---&10.93&$+0.64$&$0\fm05$&Peculiar&
\multicolumn{1}{c|}{\bf Field 6}&\multicolumn{4}{l}{{\bf HII\,1553 [S: 1]}\quad {\bf Ref: HII\,1553}}\\
309 11.27 \ \ ---&11.37&$+0.71$&$0\fm18$&\qquad --- &1508 14.14 \ ---& 14.13 &$+0.71$& --- &\qquad --- \\
322 12.58 \ \ ---&12.72&$+0.77$&$0\fm16$&\qquad --- &\#2\quad\ \ ---\quad\ ---  & 15.25 &$+0.54$& --- &\qquad ---\\
\cline{1-5} \multicolumn{1}{c|}{\bf Field 3}&
\multicolumn{4}{l||}{{\bf HII\,522 [S: 3]}\quad {\bf Ref: HII\,522}}&
1553 13.25 \ 2.31&13.28&$+1.06$& $0\fm07$& --- \\
522 12.80 \ \ 1.59&12.87&$+0.94$&$0\fm10$&Peculiar   &1554 12.56 \ ---  & 12.49 &$+0.66$& --- &\qquad --- \\
575 14.93 \ \ ---&15.26&$+1.18$& --- &\qquad --- &\#3\quad\ \ ---\quad\ ---  & 15.99 &$+0.74$& --- &\qquad \\
\cline{6-10} 598 15.88 \ \ ---&16.86&$+1.16$& --- &\qquad --- &
\multicolumn{1}{c|}{\bf Field 7}&\multicolumn{4}{l}{{\bf HII\,1776 [S: 2/3]}\quad {\bf Ref: HII\,1776}}\\
616 15.35 \ \ ---&15.88&$+1.25$& --- &\qquad --- &1776 11.60 \ 2.72&11.63 &$+0.72$& $0\fm05$&\qquad --- \\
\cline{1-5} \multicolumn{1}{c|}{\bf Field
4}&\multicolumn{4}{l||}{{\bf HII\,676 [S: 3]}\quad  {\bf Ref: HII\,738 [F: 1/2]}}&
\#4\quad\ \ ---\quad\ ---  & 17.05 &$+0.98$& --- &\qquad --- \\
625 13.71 \ \ ---&13.80&$+1.16$&$0\fm50$&\qquad --- &\#5\quad\ \ ---\quad\ ---  & 16.87 &$+1.11$& --- &\qquad --- \\
655 15.82 \ \ ---&16.28&$+1.23$& --- &\qquad --- &1844 12.29 \ ---  & 11.53 &$+0.32$& $0\fm07$&\qquad --- \\
676 14.59 \ \ 0.46&14.72&$+1.26$&$0\fm50$&\qquad --- &1900 11.43 \ ---  & 12.22 &$+1.21$& --- &\qquad --- \\
\cline{6-10} 711 16.0 \quad\,---&16.46 &$+1.15$& --- &\qquad --- &
\multicolumn{1}{c|}{\bf Field 8}&\multicolumn{4}{l}{{\bf HII\,3197 [F: 1]}\quad {\bf Ref: HII\,3197}}\\
738 13.42 \ \ 2.60&13.42&$+1.16$&$0\fm15$&\qquad --- &3167 14.23 \ ---  & 14.80 &$+0.68$& $0\fm25$&P\,$=0\fd901$\\
\cline{1-5} \multicolumn{5}{l||}{{\bf Non-Catalogued
Stars}:\hspace*{4.6mm}\vline\ \ \#4:\ \ 03\ 48\ 25.7\ \ \ +25\ 12\ 35}
&3197 13.06 \ 2.32&13.36&$+1.10$&$<0\fm03$&Very quiet\\
\multicolumn{5}{l||}{\#1:\ \ 03\ 47\ 34.1\ \ \ +23\ 45\ 06\ \
\vline\ \ \#5:\ \ 03\ 48\ 18.7\ \ \ +25\ 12\ 35}
&3198 14.89 \ --- &15.31&$+0.78$&$0\fm08$&\qquad\\
\multicolumn{5}{l||}{\#2:\ \ 03\ 47\ 39.2\ \ \ +22\ 56\ 43\ \
\vline\ \ \#6:\ \ 03\ 52\ 11.7\ \ \ +24\ 38\ 13}
&\#6\quad\ \ ---\quad\ ---  & 16.42 &$+0.85$& --- &\qquad --- \\
\multicolumn{5}{l||}{\#3:\ \ 03\ 47\ 43.4\ \ \ +22\ 57\ 07\ \
\vline\ \ \#7:\ \ 03\ 52\ 06.6\ \ \ +24\ 38\ 42}
&\#7\quad\ \ ---\quad\ ---  & 16.48 &$+0.84$& --- &\qquad --- \\
\hline
\end{tabular}
\end{center}
\end{table*}

\section{Summary Discussion}
Although the $m_{pg}$ passband in HzC does not match the CCD
$B$-band, the effective wavelengths are fairly close. An excellent
correlation between $m_{pg}$ values and the $B_i$ magnitudes, as
listed in Table~1, is displayed in Fig.~\ref{F:8}, so confirming
the data quality presented in this exercise. The systems are
related essentially by a linear transformation. Three stars which
might be considered as deviants from the correlation are
highlighted, their peculiarities being discussed earlier. The
extreme departure of HII\,282 from the $m_{pg}/B_i$ correlation
requires further investigation.

\begin{figure}
\centering
\caption {The correlation between the $m_{pg}$ values in HzC with
the $B_i$ magnitudes determined by CCD photometry is very marked.
Three labelled stars have anomalous values (see text).}
\label{F:8}
\end{figure}

From the $(B_i-V_i)\,,V_i$ values, Fig.~\ref{F:9} presents a raw
colour-magnitude diagram, without attempts to correct for
interstellar extinction, this generally being very small (see
Crawford \& Perry, \cite{CP}). The diagram reveals the trend of
the target stars, following a smooth extension of the cluster main
sequence. Inspection shows two stars (HII\,1900 and HII\,272) well
above this line. The enigmatic spread below the main sequence
trend is very apparent. At $(B_i-V_i)=+0.8$, the apparent
luminosity covers $\sim 5$ magnitudes. Such dispersion cannot
result from large variations of interstellar extinction from star
to star; simple consideration of the vector rule that $A_V\approx
3E(B-V)$ does not resolve the problem. Although some parts of the
cluster are contaminated by interstellar extinction, the effects
are far too small to provide the explanation.

\begin{figure}
\centering
\caption {The HR diagram based on raw $V_i/(B_i-V_i)$ measurements
covering solar-type Pleiads. The 8 target stars selected to
investigate the $A_{\rm Li}$/surface activity connection are
highlighted with heavy dots and form an extension of the main
sequence of the brighter stars in the cluster. The enigmatic
spread of the HR diagram is very apparent, with a few stars
(HII\,1990 and HII\,272) above the main sequence, suggesting
unresolved duplicity, but with many fainter stars below the
sequence trend. Other assigned numbers correspond to stars with
noted variability. Non-numbered stars appear photometrically
quiet, but because of their faintness and poorer measurement
accuracies, this generalisation  may be imperfect.} \label{F:9}
\end{figure}

Two monitored stars (HII\,282 and HII\,522) appear to display
fadings superimposed on the rotational modulation of photospheric
disturbances. Their positions in the colour/brightness diagram
suggests they might be photometric binaries. One explanation of
the fading behaviour is that they are short-period eclipsing
binaries with a dwarf companion. The signature is also similar in
character to the behaviour of HAEBE stars. For the latter, the
consensus model is based on orbiting disk-like structures with
irregularities in dust densities. If they do possess circumstellar
dust, any extinction must be fairly neutral. To reiterate,
HII\,282 certainly deserves follow-up studies with multi-colour
photometry and polarimetry.

Four stars have had periods assigned in Table 1. The value of
$0\fd777$ for HII\,1532 is in excellent agreement with a previous
determination by Krishnamurthi et al. (\cite{Ketal}). For the
other stars, the periods are tentative and further observations
are desirable to confirm their values.

Finally, in respect of the purpose of this study, of the nine
stars with catalogued $A_{\rm Li}$ measures, comparison with their
observed $V_i$ variation, confirms that there is a relation,
although there are deviants degrading the correlation. The star
with by far the largest variability of the monitored sample
(HII\,676), was not expected to be all that active from its
$A_{\rm Li}$ value. Also HII\,3197 provides an anomalous result.
According to its value of $A_{\rm Li}$, it would have been
expected to display photometric changes $\sim 0\fm2$, whereas any
activity was at least a factor of 10 smaller. Other factors
affecting the $A_{\rm Li}$ values of individual stars must
therefore be present. One fact is inescapable, however; the
majority of solar-type stars in the cluster display photometric
variability with brightness changes of $0\fm01$ or larger,
representing the effects of the rotational modulation of
photospheric disturbances. Their activity suggests that, with a
good programme of high quality photometry, extension of the data
base for measured periods of Pleiads should be guaranteed.

\begin{acknowledgements} ECM was supported as a summer student by
the Department of Physics and Astronomy of the University of
Glasgow and SO received maintenance as a PPARC research student.

\end{acknowledgements}

\end{document}